# Strain driven emergence of topological non-triviality in YPdBi thin films

Vishal Bhardwaj[1], Anupam Bhattacharya[2], Shivangi Srivastava[1], Vladimir V. Khovaylo[3], Jhuma Sannigrahi[4], Niladri Banerjee[4+], Brajesh K. Mani[1] and Ratnamala Chatterjee[1*]

[1]Department of Physics, Indian Institute of Technology Delhi, India
[2]Department of Mechanical Engineering, Indian Institute of Technology Delhi, India
[3]National University of Science and Technology "MISiS," Moscow 119049, Russia
[4]Department of Physics, Loughborough University, Loughborough, LE11 3TU, United Kingdom

[*]ratnamalac@gmail.com

[+]N.Banerjee@lboro.ac.uk

Half-Heusler compounds exhibit a remarkable variety of emergent properties such as heavy-fermion behaviour, unconventional superconductivity and magnetism. Several of these compounds have been predicted to host topologically non-trivial electronic structures. Remarkably, recent theoretical studies have indicated the possibility to induce non-trivial topological surface states in an otherwise trivial half-Heusler system by strain engineering. Here, using magneto-transport measurements and first principles DFT-based simulations, we demonstrate topological surface states on strained [110] oriented thin films of YPdBi grown on (100) MgO. These topological surface states arise in an otherwise trivial semi-metal purely driven by strain. Furthermore, we observe the onset of superconductivity in these strained films highlighting the possibility of engineering a topological superconducting state. Our results demonstrate the critical role played by strain in engineering novel topological states in thin film systems for developing next-generation spintronic devices.

# Introduction

The conventional scheme of classification of materials into metals and insulators has been recently challenged by the discovery of topological insulators[1–3]. This new state of quantum matter has a fully gapped insulating state in the bulk and topologically protected gapless surface or edge states. Intriguingly, they show low-energy electronic excitations that resemble the Dirac[4] and Weyl[5] fermions enabling us to explore the physics of these elusive particles in high energy physics using much simpler condensed matter experiments[6]. These relativistic fermions lead to several exotic transport characteristics – from chiral anomaly to the intrinsic anomalous Hall effect. Recently, semimetals has also been included in this scheme of topological classification [3,6] further widening the available material systems displaying these remarkable properties. The non-trivial topology of bulk bands in these semimetals leads to the topologically protected Fermi arcs and drumhead surface states[6].

The half-Heusler alloy family consists of a large collection of semimetals that are predicted to exist in either topologically trivial or non-trivial state[7–9]. The key advantage in Heusler systems is the ability to precisely tune the bandgap and band inversion strength by altering the chemical composition. Even more interestingly, recent theoretical studies have indicated the possibility to engineer non-trivial topological states in an otherwise trivial half-Heusler system by strain engineering. While these non-trivial surface states can be detected in ultra-pure samples by angle resolved photoemission spectroscopy and other surface sensitive techniques[10–13], transport measurements provide a relatively easy way to detect the surface states[6,14–18]. This is possible since the surface states are topologically protected and conserve time reversal symmetry which can be broken by an applied external magnetic field providing a pathway to detect them using transport measurements[3,6,16,19]. The distinct signatures in transport measurements are extremely large magneto-resistance, sharp cusp in magneto-resistance around low magnetic field, negative longitudinal magneto-resistance and quantum oscillations in the presence of high magnetic field. These signatures originate from the carriers having light effective mass, high mobility, chiral anomaly, and nontrivial Berry phase associated with them. The nontrivial Berry phase is a more convincing parameter to ascertain topological materials using magneto-transport studies because it stems from the time reversal symmetry protection of the carrier wave functions. The Berry phase can be extracted by measuring temperature dependent magneto-resistance data and analyzing weak-antilocalization effect in low magnetic field and Shubnikov de-Hass oscillations in high field magneto-resistance data separately [3,6,20].

YPdBi is a well-established topologically *trivial* half-Heusler system[7–9] based on nuclear magnetic resonance[21–24], electron spin resonance[25,26] and from bulk band structure calculated using density functional theory (DFT)[22]. Although bulk unstrained YPdBi is topologically trivial, interestingly it was theoretically predicted to undergo a strain-driven transition to a non-trivial state[9]. This strain in bulk single crystals could directly result from application of external pressure or for thin film systems, could be engineered by growing epitaxial/oriented films on suitable substrates. We note that although Wang *et. al.* reported SdH oscillations and large linear MR in (100) YPdBi single crystals, the origin of these oscillations was entirely attributed to the highly mobile 3D bulk electron carriers and no direct evidence of non-triviality was observed[27].

In this study, we report the magneto-transport measurements on (110) oriented strained YPdBi thin films (~30nm) grown on single crystal MgO (100) substrates. By analyzing the magneto-transport data we demonstrate the non-trivial nature of the surface states of these oriented YPdBi films. We also observe a sharp resistance drop below 1.25 K which possibly indicates the onset of superconductivity. To support our experimental results, we perform DFT based first-principles calculations to show a topologically trivial to non-trivial transition driven by a strained YPdBi thin film. A distinct advantage in thin film systems is the large surface/volume ratio making it easier to detect the effect of the surface states on transport properties with a reduced contribution from the bulk.

## Results

**Structural characterizations:** Figure 1(a) shows the powder XRD pattern and Rietveld refinement of bulk YPdBI sample. The Rietveld refinement of powder XRD data confirms the $C1_b$ crystal structure (F$\bar{4}$3m space group) of the sample with lattice constant ~ 6.638 Å[27,28]. Figure 1(b) shows the Gonio mode XRD pattern of YPdBi thin film and the observation of Bragg reflections corresponding to (220) and (440) planes indicate the (110) oriented growth of the film with the expected $C1_b$ structure and lattice constant ~ 6.845 Å. This indicates about ~ 3.12 % strained lattice of YPdBi thin films. A narrow full width at half maximum (FWHM) of about ~

0.19° obtained from the rocking curve (ω -2θ) scan around (220) plane (inset of the Fig. 1(b)) indicates the high crystallinity and (110) orientation growth of YPdBi thin film. Using specular X-ray reflectivity (XRR) (ω -2θ) scans we calibrate the deposition rate and estimate the interfacial roughness of our films. Figure 1(c) shows the experimental and fitted XRR spectra for YPdBi ~ 30 nm thin films on a Ta ~ 5 nm layer with interface roughness ~ 1.2 nm. Insets of Fig. 1(c) show the $C1_b$ crystal structure of YPdBi and schematic of Ta (5nm)/ YPdBi (30nm) bilayer stack. Using AFM we analyse the film topography and estimate the surface roughness ~ 1 nm from the AFM image as shown in Fig. 1(d) with the granular growth of the film shown in the tilted 3D AFM image inset.

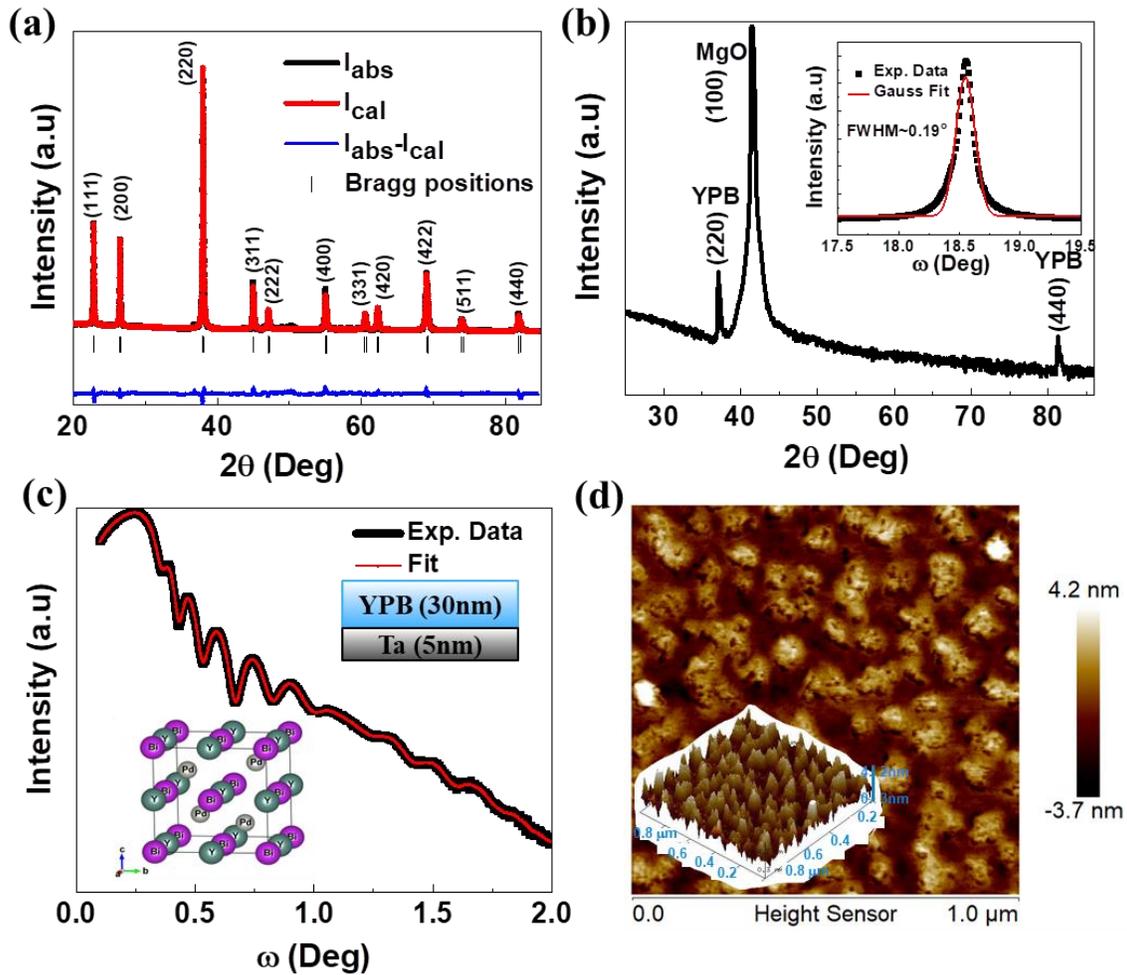

**Figure 1| Structural characterizations of YPdBi :** (a) Rietveld refinement of YPdBi powder XRD pattern. (b) Gonio mode XRD pattern of YPdBi thin film, inset shows rocking curve scan of (220) peak. (c) X-ray reflectivity scan of Ta (5nm)/YPdBi (30 nm) and corresponding fitting

curve in red color. Insets show the $C1_b$ lattice structure of YPdBi and schematic stacking of Ta/YPdBi thin films. (d) Topographical AFM image, inset shows corresponding tilted 3D image.

Magnetic field ($H$) – and temperature-dependent resistance ($R$) measurement in a four-point current-biased configuration were performed on unpatterned samples of dimensions 3 mm × 10 mm[15,17]. Figure 2(a) shows the temperature-dependent resistivity ($\rho_{xx}$) data of YPdBi thin film in the temperature range of 1.9 K – 350 K. The data show a slower rise of resistivity with decrease in temperature as expected from a semi-metal in the temperature range 2.2 K ≤ T ≤ 350 K. Interestingly, there is a sharp down-turn in the resistivity below ~ 2.2 K with a corresponding transition width (ΔT) = 0.95 K ($T_C$ ~ 1.25 K), see inset of Fig. 2(a). This sharp drop in resistivity could be linked to the onset of superconductivity.

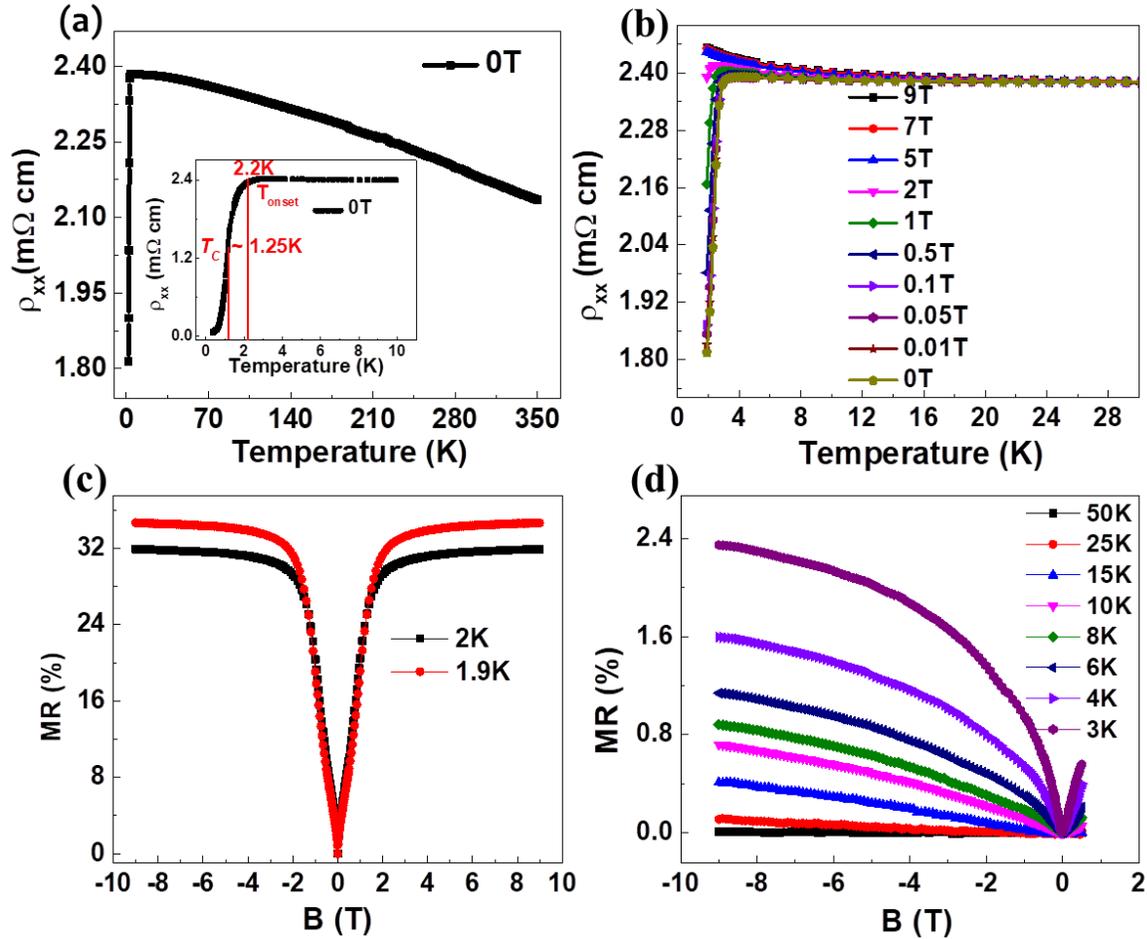

**Figure 2 | Electrical and magneto-transport properties of YPdBi :** (a) $\rho_{xx}$ as a function of temperature in range 1.9 K ≤ T ≤ 300K , inset shows $\rho_{xx}$ measured from 0.38 K to 10 K . **(b)**

$\rho_{xx}$ vs temperature measured in presence of magnetic field from 0 T to 9 T in temperature range 1.9K-50K. **(c)** MR data at temperatures 1.9 K and 2 K. **(d)** MR data in the temperature range 3 K ≤ T ≤ 50 K.

**Electrical and magneto-Transport properties:** To gain more insight into the possible onset of superconductivity in YPdBi at low temperatures, we measured this drop in $\rho_{xx}$ in presence of an external magnetic field of increasing strength from 0 T to 9 T applied perpendicular to the film plane. The data, shown in the Fig. 2(b), clearly indicates that this resistance drop completely disappears for magnetic fields ≥5 T. These measurements were limited by the base temperature of the cryostat at 1.9 K. However, we measured the zero-field resistivity in a separate system with a base temperature of 0.38 K and the data is plotted in the inset of Figure 2(a) which clearly shows a sharp drop in resistivity by almost two orders of magnitude from 2.2 K down to 0.38 K. The residual resistivity value at the base temperature of the cryostat is 0.05 mΩ-cm.

A series of $R-H$ measurements were performed at various temperatures with a perpendicularly applied field and the corresponding fractional resistance change (called magneto-resistance or MR) calculated using the formula $[\{R(H) - R(0)\}/R(0)] \times 100$ where $R(H)$ and $R(0)$ are the resistances of the sample at magnetic field H and zero respectively. Figure 2(c) shows the MR% of YPdBi recorded at 1.9 K and 2 K with maximum MR of ~ 35% observed at 1.9 K. The full dependence of MR above this temperature is showed in Figure 2(d) up to a maximum temperature of 50 K. Although large MR is observed for temperatures 1.9 K and 2 K (Figure 2(c)) where the film shows metallic characteristics, the MR decreases to ~ 2.4% at 3 K and vanishes completely at 50 K as seen in Fig. 2(d). The prominent cusp in the MR behaviour at low fields is indicative of weak anti-localization (WAL) – a signature of the conductive surface states of topological insulators[29–31] and topological semimetals[18,20,32,33] arising from quantum interference corrections of the diffusive transport. We fit our MR results around low magnetic field region (±0.5T) to the well-known Hikami-Larkin-Nagoka (HLN) model, which describes the conductivity of a two dimensional electron systems with strong spin-orbit interactions under an applied magnetic field[34]:

$$\Delta G_{xx} = \frac{-\alpha e^2}{2\pi^2 \hbar}\left[\ln\frac{B_\varphi}{B} - \Psi\left(\frac{1}{2}+\frac{B_\varphi}{B}\right)\right] \quad (1)$$

where $B_\varphi = \frac{\hbar}{4eL_\varphi^2}$ and $\hbar$, $e$ and $L_\Phi$ are the reduced Planck's constant, the charge of the electron and the phase coherence length of the carriers, respectively. The function $\psi$ is the digamma function and the parameter $\alpha$ defines the number of coherent channels, i.e., $\alpha$ = -0.50 and -1 for single and two coherent channels, respectively.

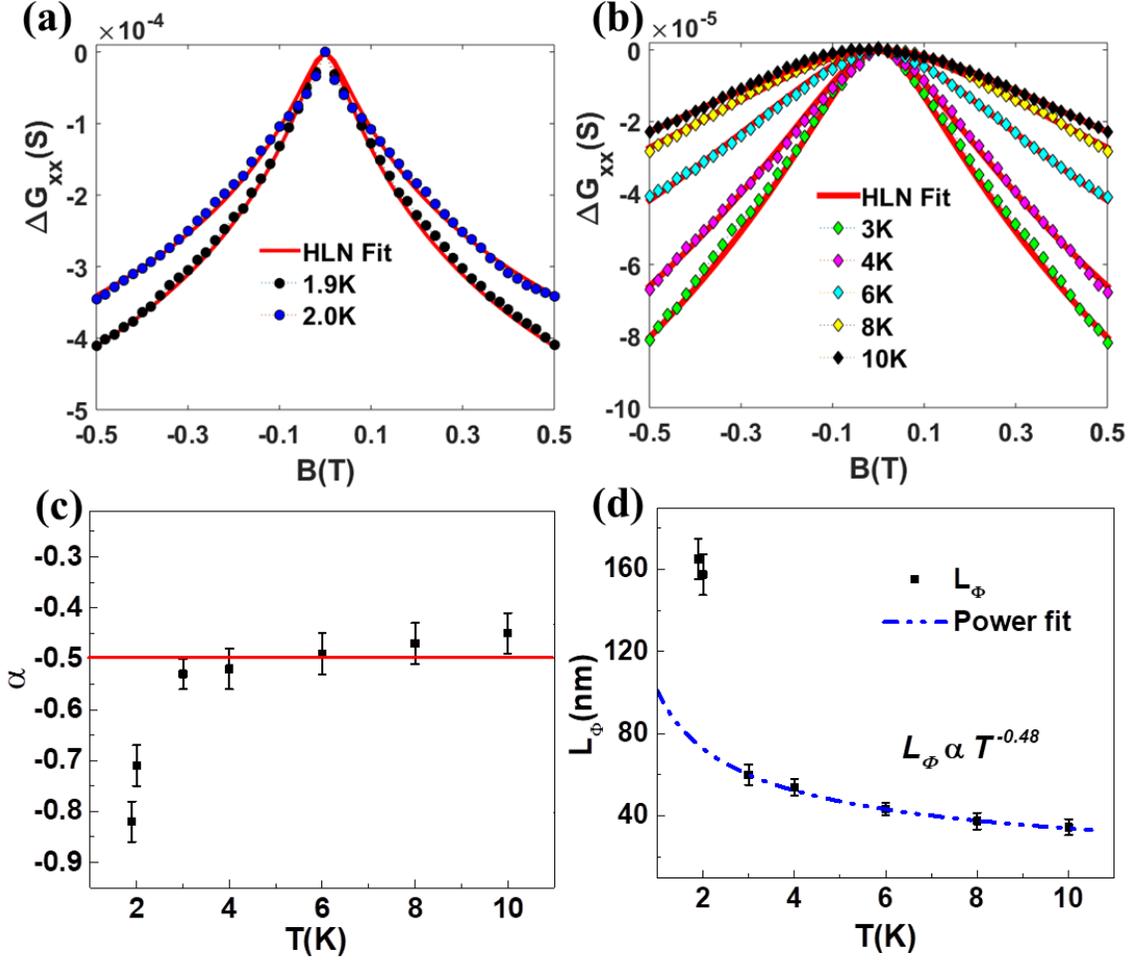

**Figure 3| Weak antilocalization effect in YPdBi:** **(a)** The HLN fit of magneto-conductance $\Delta G_{xx}$ data at 1.9 K and 2 K. **(b)** HLN fitting in range 3 K ≤ T ≤ 10 K. **(c)** Variation of $\alpha$ with temperature. **(d)** Variation of $L_\Phi$ with temperature.

Figure 3(a) (for T = 1.9 K and 2 K) and Fig. 3(b) (3 K ≤ T ≤ 10 K) show the HLN model fitted longitudinal magneto-conductance $(\Delta G_{xx} = G_{xx}(B) - G_{xx}(0))$ data in a strong spin-orbit interaction regime, i.e., when the inelastic scattering time $(\tau_\phi)$ >> spin-orbit scattering time $(\tau_{SO})$ and elastic scattering time $(\tau_e))^{34-36}$. The variation of the HLN fitted parameters, the pre-factor ($\alpha$) and the phase coherence length ($L_\Phi$) with temperature (1.9 K ≤ T ≤ 10 K) is shown in the Figs. 3(c) and 3(d), respectively. $\alpha$ estimated from fitting the $\Delta G_{xx}$ data is ~ -0.5 in the temperature range $3\ K \leq T \leq 10\ K$, but drops to -0.71 at 1.9 K and -0.82 at 2 K. The value of -0.5 indicates a single surface conducting channel while lower values of -0.71 and -0.82 indicates contributions from parallel bulk conducting channels. This contribution can either come from intermixing with bulk conducting channels or could also be induced by the onset of superconductivity below 2.2 K (Fig. 2(a) inset). Further evidence of the 2D nature of the system comes from fitting the temperature dependence of $L_\Phi$ as shown in Figure 3(d). The power-law dependence with an exponent of -0.48 is remarkably close to -0.5 for a 2D system. A very high value of $L_\Phi$ is observed at 1.9 K and 2 K as 162nm and 158nm respectively. The $L_\Phi$ decreases to ~ 40 nm at higher temperatures (10K) which can be attributed to increased electron-phonon scattering with increase in temperature.

To summarize, the data in Fig. 2 shows a typical semi-metallic curve in temperature range 2.2 K to 350 K and an onset of superconductivity below 2.2 K with $T_C$ ~ 1.25 K. The sharp drop in resistivity disappears in presence of a magnetic field H ≥ 5T. A positive saturating MR ~35% is observed at 1.9K (Fig. 2(c)) and the MR data shows WAL effect around low magnetic fields - a key signature of topological non-triviality. Figure 3 summarizes the fitting of the WAL to well-known HLN equation indicate the observation of 2D-WAL effect with $\alpha$ ~ -1/2, $L_\Phi \propto T^{-0.48}$ above the transition ($T_C$ ~ 1.25 K) and contribution of bulk atoms in the conductance below this transition ( with $\alpha$ =−0.71 at 1.9 K and −0.82 at 2 K ). We will discuss the contribution of surface and bulk atoms to the conductance using DFT in next sections.

**Quantum oscillations:** The quantitative information about the dimensionality of the Fermi surface and non-triviality of the material system calculated from the Berry phase can be extracted from the quantum oscillations of the conductivity – the Shubnikov-de Haas (SdH) oscillations[20,33,37–39]. The SdH oscillations originate in the presence of high magnetic field where the density of states of a crystalline solid gets modulated periodically as a function of the magnetic field ($B$) due to the Landau quantization[6,37].

Figure 4(a) shows the SdH oscillations, periodic in $1/B$, extracted from the high magnetic field MR data (after the background subtraction) in the temperature range 1.9 K ≤ T ≤ 10 K. The clear SdH oscillations are observed up to 15 K, above which the oscillatory signal is masked by the background noise. The inset of the Fig. 4(a) shows the fast Fourier transformed (FFT) spectra of SdH oscillations at various temperatures and the extracted single frequency of oscillations is $f_{SdH} \sim 34$ T. This frequency is directly related to the Fermi-surface cross section $A_F$ normal to the magnetic field which can be estimated using the Onsager relation $A_F = (2\pi e(f_{SdH}/\hbar))$. This gives a Fermi wave vector ($k_f$) and sheet carrier concentration ($n_s = k_f^2/4\pi$) of 0.032 Å$^{-1}$ and 8.21 ×10$^{11}$ cm$^{-2}$ respectively. We note that if these oscillations originated from the bulk, to obtain a similar Fermi wave vector would require a significantly higher carrier density.

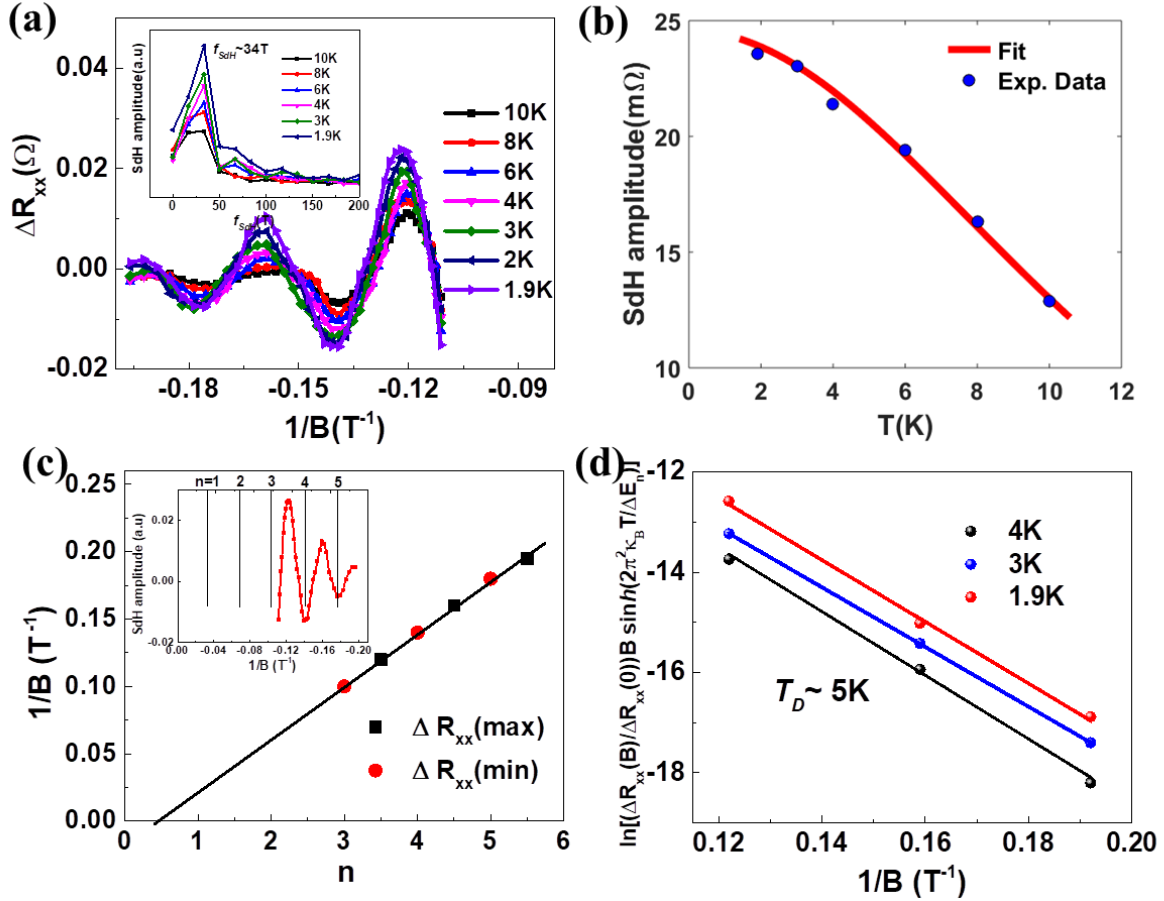

**Figure 4| Lifshitz-Kosevich analysis of SdH oscillations: (a)** The SdH oscillations in resistance data after background subtraction as a function of 1/B in temperature range 1.9 K– 10 K. Inset shows the corresponding FFT spectra. **(b)** Standard L-K fit of temperature dependent SdH amplitudes. **(c)** The Landau level fan diagram of SdH maximas and minimas. Inset shows the Landau level assignment to SdH data at 1.9K. **(d)** Dingle plots at 1.9 K, 3 K and 4 K.

We estimate the cyclotron effective mass of the carriers by analyzing damping of the oscillations with increase in temperature which is fitted with the thermal damping term of the standard Lifshitz-Kosevich expression[40,41]: $\frac{2\pi^2 \kappa_B T/\Delta E_n(B)}{\sinh(2\pi^2 \kappa_B T/\Delta E_n(B))}$ (Fig. 4(b)) Here, $\kappa_B$ is Boltzmann's constant and $\Delta E_n(B)$ is the fitting parameter which is related to the cyclotron effective mass (m*) of the carriers as $\Delta E_n(B) = \hbar eB/m^*$. We obtain an effective mass $m^* = 0.12 m_e$, where $m_e$ is the free electron mass. The parameter $\Delta E_n(B)$ accounts for the splitting of the Landau levels ($n$) in presence of the magnetic field, $B$. The minima and maxima of oscillations in $\Delta R_{xx}$ correspond to the Landau levels $n$ and $n + \frac{1}{2}$, respectively, as shown in the inset of the Fig. 4(c).

In normal metals the cyclotron orbits of electrons result in a zero Berry phase (= $2\pi\beta$ with $\beta = 0$), whereas $\beta = 1/2$ results in a $\pi$ Berry phase for a system with linear dispersion at the degenerate (Dirac) point[42–44]. We estimate the Berry phase from the Landau level fan diagram, where $n$ data points are plotted as a function of inverse magnetic field and least square fitted to a straight line that result in intercept $\beta$ on $n$ axis (shown in the Fig. 4(c)). This gives $\beta$ = 0.46±0.1 and Berry phase = 0.92±0.1 from $n$ axis intercept of linear fitting to $(\Delta R_{xx})_{max}$ and $(\Delta R_{xx})_{min}$ data points. Assuming electrons are Dirac type, we estimate the Fermi velocity by using $v_F = \hbar k_f/m^*$ and position of the Fermi level from linear bands crossing ($E_F^S = m^* v_F^2$) as ~ $3.21 \times 10^5$ ms$^{-1}$ and 68 meV respectively. These values are of the same order as reported for topologically non-trivial LuPdBi topological semimetal[33]. The transport lifetime and mean free path of the carriers are estimated using Dingle plot of SdH oscillations observed at a temperature T by extracting the slope of $\ln(\Delta R_{xx}(B)/\Delta R_{xx}(0))B\sinh(2\pi^2\kappa_B T/\Delta E_n)$ vs $1/B$ plot[15,17,27]. Figure 4(d) shows the Dingle plots at 1.9 K, 3 K and 4 K, the slope of linear fitting to the data points gives Dingle temperature $T_D$ ~ 5 K. The transport lifetime ($\tau$) and mean free path ($l$) of carriers are estimated using the relations, $\tau = \hbar/(2\pi T_D)$ and $l = V_F \tau$ as ~ $2.43 \times 10^{-13}$ s and 78 nm, respectively. The carrier mobility ($\mu_s = e\tau/m^*$) ~ 3694 cm$^{-2}$V$^{-1}$s$^{-1}$ is estimated for strained YPdBi thin films, which is quite high among other non-trivial half Heusler alloys like LuPdBi (2100 cm$^{-2}$V$^{-1}$s$^{-1}$)[33], ErPdBi (1035 cm$^{-2}$V$^{-1}$s$^{-1}$)[14], DyPdBi (1780 cm$^{-2}$V$^{-1}$s$^{-1}$)[17] and YPtBi(1486 cm$^{-2}$V$^{-1}$s$^{-1}$)[18]. All the parameters extracted from the different fits are listed in the Table 1

**Electronic structure of YPdBi bulk and strained thin films:** The experimental results reported so far strongly suggest topologically non-trivial band structure of strained YPdBi thin films. In order to confirm this, we perform DFT-based first principles calculations. The crystal structure of YPdBi is rhombohedral associated with the space group $F\bar{4}3m$. The unit cell contains three atoms where atoms Y, Pd and Bi occupy the (0.5, 0.5, 0.5), (0.25, 0.25, 0.25) and (0.0, 0.0, 0.0) Wyckoff positions, respectively. We calculate the equilibrium lattice constant, which is obtained by minimizing total energy with respect to the lattice parameter. The calculated lattice constant of 6.570 Å agrees well with the measured value, 6.638 Å. The difference between the two results can be attributed to the use of LDA pseudopotentials, which are known for underestimating lattice parameters by approximately 1% [45].

We start by first reproducing the electronic band structure of the unstrained YPdBi bulk (Fig. 5(a), upper panel) from Ref 9. As expected, the band dispersion of unstrained YPdBi lattice shows a doubly degenerate $\Gamma_6$ lying above the quadruply degenerate $\Gamma_8$ indicating a topologically trivial band structure[46], in agreement with the previously reported theoretical and experimental studies on YPdBi[21–23]. Next, we apply a tensile strain of 3.12% along the [001] direction corresponding to a lattice constant of 6.774 Å which is similar to the lattice constant we obtain for our strained films from the XRD data shown in Fig. 1(a). The calculated band structure is shown in the lower panel of Fig. 5(a); there is a clear band inversion at the $\Gamma$ point with $\Gamma_6$ falling below $\Gamma_8$ in the E-k diagram suggesting topological non-triviality in the strained lattice of YPdBi. Our DFT results, showing a strong topological nature for the strained YPdBi lattice, agrees with the conclusions from our experimental magneto-transport results and previous calculations with similar, albeit a smaller, strain[9]. A closer examination of the strained band structure from Fig. 5(a) (lower panel) shows that the conduction and valence bands YPdBi touch at $\Gamma$ point and obey a quadratic dispersion relation. We estimate the inertial effective mass ($m^{**}$) of the electron at the center of the Brillouin zone to be $0.048m_e$ by calculating the curvature of the valence bands near the $\Gamma$ point. The very low effective electron mass results in the high mobility of electrons at the $\Gamma$ point in the case of strained bulk YPdBi, which reflected through observation of SdH oscillations in MR experiments.

Although we see clear evidence of band inversion in the strained YPdBi lattice, the calculations do not capture the finite extent of our thin films which can introduce further modifications of the band structure. To simulate our strained YPdBi thin film, we create a semi-infinite slab made of 20 atomic layers with strained lattice constant of 6.774 Å, as shown in the Fig. 5(b). A vacuum of approximately 20 Å is placed at the end of the slab to simulate the semi-infinite boundary condition of the film. Out of the 20 atomic layers of the slab, two outer layers each from top and bottom are referred as the surface layers and the rest 16 layers are referred as bulk layers. To assess the contribution of the surface states to electronic transport through the film, we calculate the k-resolved density of states of the slab as shown in Fig. 5(c). Electronic density from the surface layers are shown in red and that from the bulk layers is shown in cyan color.

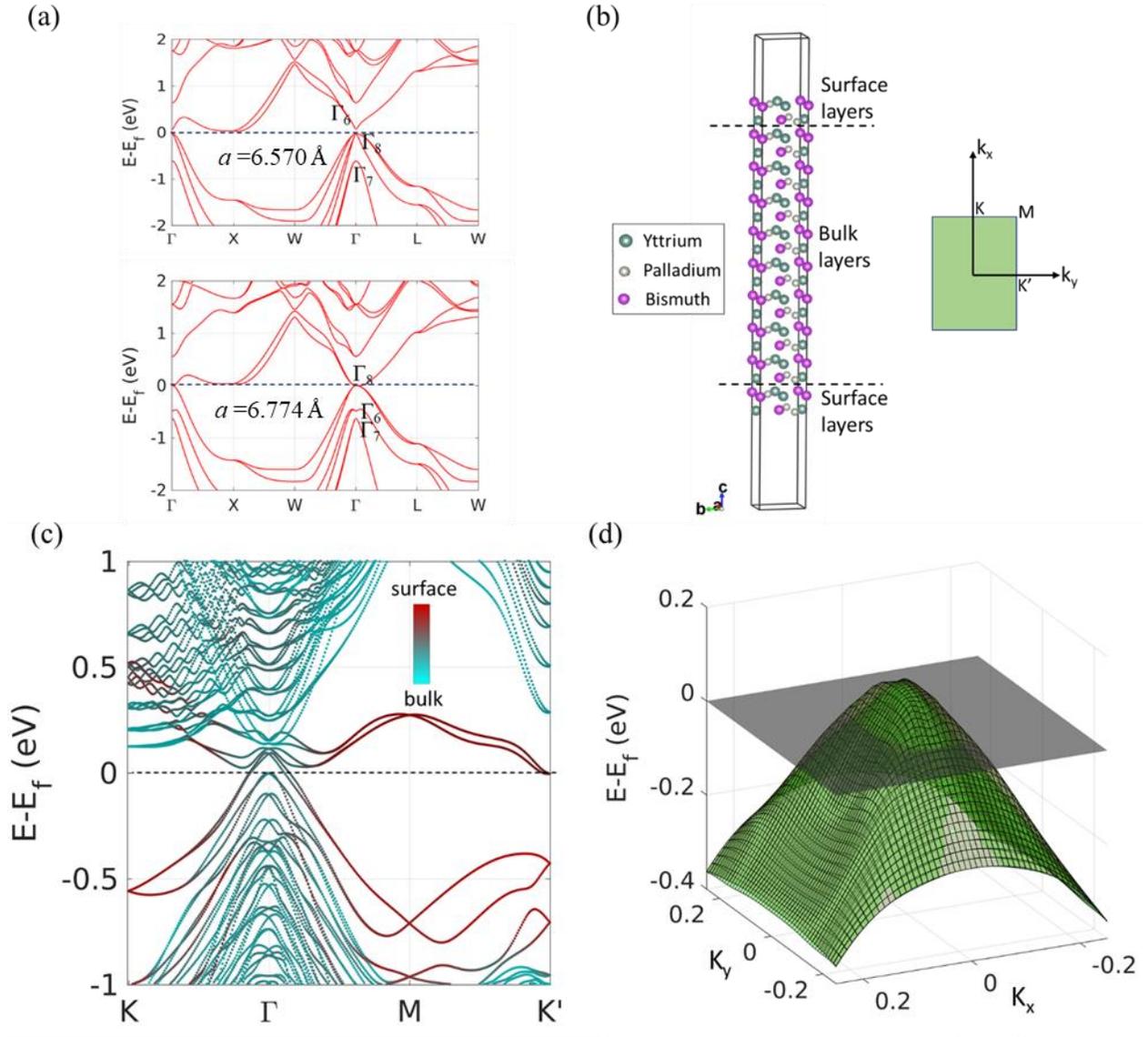

**Figure 5| DFT-based first principles calculations: (a)** Electronic band structure calculated with equilibrium lattice constant 6.570 Å and strained lattice constant 6.774 Å. **(b)** Lattice structure of semi-infinite (110) oriented slab of YPdBi with vacuum on both sides. **(c)** The K-resolved

density of states showing contribution from surface atoms. Red color represents the contribution of atoms from the surface layers and cyan color represents the contribution from bulk layers. **(d)** Energy dispersion of two bands crossing the Fermi energy.

Similar to the strained bulk lattice, a parabolic band dispersion is seen at the $\Gamma$ point (Fig. 5(c)) and the two bands that cross the Fermi energy near $\Gamma$ point, create the Fermi pockets. However, in contrast to the bulk band dispersion, we observe a lower Fermi energy and surface atom contributed states (red) just above and below the Fermi energy. Strikingly, we see that around $\Gamma$ point, these two bands have contributions from both the surface atoms as well as bulk of the slab. In order to precisely calculate the inertial effective mass $m^{**}$ of the electrons originating from this Fermi pocket, we separated out these two bands (see Fig. 5(d)) and calculated the curvature at the local maxima ($\Gamma$ point) of these bands. The calculated $m^{**}$ tensors for these two bands at $\Gamma$ point are

$$(m^{**})_{band1} = \begin{vmatrix} -0.0028 & 0.0000 \\ 0.0000 & -0.0368 \end{vmatrix} \text{ and} \tag{2}$$

$$(m^{**})_{band2} = \begin{vmatrix} -0.0327 & 0.0000 \\ 0.0000 & -0.0416 \end{vmatrix} \tag{3}$$

normalized with respect to the rest mass of the electron, $m_e$. Such low values of $m^{**}$ give rise to the high mobility of electrons and SdH oscillations in (110) oriented YPdBi thin film. Identifying these Fermi pockets as the origin of very mobile electrons on (110) surface, we reach a value of $E_f^s$=88 meV, which is of the same order of magnitude with experimentally observed value of $E_F^s$, as given in Table 1. We note that since the dispersion shown around $\Gamma$ in Fig. 5(c) is approximately parabolic, effective mass calculated through curvature is reasonably accurate. However, the scatterings at defects and impurities are expected to affect the experimentally obtained mobility and effective mass of the electrons originated from the electronic pockets. This could possibly explain the lower values of calculated $m^{**}$ compared to the experimentally measured value (Table 1).

**Table 1. Parameters extracted from the fit of SdH data:**

| $f_{SdH}$ (T) | $n_s$ ($10^{11}$cm$^{-2}$) | $m^*$ ($m_e$) | $k_f$ (Å$^{-1}$) | $V_F$ ($10^5$ms$^{-1}$) | $E_F^s$ (meV) | $\tau$ ($10^{-13}$s) | $l$ (nm) | $\mu_s$ (cm$^2$V$^{-1}$s$^{-1}$) |
|---|---|---|---|---|---|---|---|---|
| 34 | 8.21 | 0.12 | 0.032 | 3.21 | ~68 | 2.43 | ~78 | ~3694 |

## Discussion:

We summarize the conclusions from our measurements and DFT calculations: The WAL in magneto-transport (Fig. 2(c)) reflected by the cusp in MR at low fields (negative magneto-conductance) arises from the destructive interference of spin ½ carrier wave functions with an associated $\pi$ Berry phase. Analyzing this data within the HLN framework shows that in the temperature range of 3 K ≤ T ≤ 10 K, $\alpha$ is approximately -0.5 indicating that the dominant transport is through the surface states. Furthermore, in the same temperature range, the $L_\Phi$ shows a power-law dependence with an exponent of -0.48 which indicates the two-dimensional nature of the transport. The observation of this two-dimensional WAL effect is strongly indicative of transport through topologically protected surface states. This is remarkable considering the fact that YPdBi is a semi-metal and naturally a contribution from the bulk conduction channels is expected. However, our data strongly suggest dominant contribution of surface states in transport in the temperature range 3 K ≤ T ≤ 10 K. We independently confirm a Berry phase of ~ 0.92 $\pi$ from analysing the SdH oscillations that further shows a large Fermi wavelength; this large Fermi wavelength cannot be reconciled with a bulk Fermi surface for that would require a significantly higher carrier density.

Our DFT calculations show the effect of a tensile strain similar to the ones in our thin films that drives the bulk YPdBi electronic band structure into a non-trivial state by establishing negative band inversion strength around the $\Gamma$ point. Further insight is provided from simulations of the semi-infinite slab of (110) oriented strained YPdBi which reveals the existence of Fermi pockets with carriers having very low $m^{**} = 0.048 m_e$. The observation of SdH oscillations and very small $m^*$(~0.12$m_e$) with high $\mu_s$ (~3694 $cm^{-2}V^{-1}s^{-1}$) can be explained by accounting for these electrons originating from the Fermi pockets.

We now comment on the possible onset of superconductivity observed below 2.2 K with a transition temperature of ~ 1.25 K. While it is difficult to comment on the nature of superconductivity from resistance measurements alone, but the possibility of topological superconductivity or unconventional pairing cannot be entirely ruled out. Moreover, the non-centrosymmetric nature of the crystal should mix singlet and triplet pairing states and previous reports have indicated the presence of odd parity states in YPdBi from upper critical field measurements [47] albeit in a bulk topologically trivial crystal. Recently, Radamanesh *et. al.*[48]

found that even for the *trivial* YPdBi single crystals the temperature dependent magnetic field penetration depth follows a power law instead of expected exponential behavior for BCS superconductors, indicating the unconventional nature of superconductivity[48]. Our resistivity data shows one key difference with the bulk superconductivity in YPdBi single crystals: the onset temperature of superconductivity is at least 0.5 K higher in thin films compared to single crystals and the transition width in thin films is also significantly larger. Although the larger transition width could simply arise from a distribution of transition temperatures in different regions of the film, but an enhancement of the onset temperature cannot be easily explained. Additionally, we note that the sheet carrier concentration for our strained YPdBi thin films is $\sim 8 \times 10^{11}$cm$^{-2}$ and is similar to that of YPtBi single crystals ($n_{2D} \sim 6 \times 10^{11}$cm$^{-2}$) reported in Ref 18. Interestingly, relativistic DFT calculations[49] show that such a small carrier concentration cannot explain the observed $T_C$ of ~0.8K in YPtBi and related half Heusler systems.

In conclusion, we have demonstrated the appearance of non-trivial topological surface states in strained [110] oriented thin films of a trivial semi-metal YPdBi. Additionally, we also observe the appearance of superconductivity in these topologically non-trivial strained films with an enhanced transition temperature compared to its bulk value. While our DFT calculations indicate that these surface states arise from a strain-driven negative band inversion around the Γ point, the exact nature of the superconductivity in these strained films require further investigation. The realization of topological surface states in half-Heusler thin films through strain engineering is important both fundamentally and from the point of view of applications. While our results open up the possibility to study a wide-range of exotic topology-driven phenomena including chiral anomaly, large magnetoresistance and unconventional superconductivity in thin film systems, it also raises the intriguing possibility to design next-generation devices for spintronic and quantum computation applications.

## Methods:

**Sample preparation:** The YPdBi thin films are grown using pulsed laser deposition system with KrF excimer pulsed laser source (λ= 248 nm). The energy density of source is about ~ 1.1 J/cm$^2$ and chamber base pressure ~ 3 ×10$^{-7}$ mbar. Laser pulses are bombarded on 1" YPdBi target, prepared using RF induction melting method[17]. Thin films are grown on MgO (100) substrate with ~ 5 nm Ta seed layer at 270 $^o$C insitu substrate temperatures.

**Experimental characterization:** The crystal structure and film thickness are determined using Cu K$_\alpha$ Panalytical X'pert highscore Diffractogram by using X-ray diffraction and x-ray reflectivity techniques. The surface topography is studied using Bruker Dimension 3100 atomic force microscope (AFM). The electrical contacts for transport measurements are prepared using copper wires and cured with silver paste. The magneto-transport properties of thin films with dimensions 3 × 10 mm are studied using Quantum Design 9T PPMS system in the temperature range from 1.9 K to 300 K. The mK transport studies are carried out using Cryogen Free Measurement System from Cryogenic Limited.

**First-principles simulations:** Density functional theory based first-principles calculations using the projector augmented wave (PAW)[50,51] method as implemented in Vienna ab-initio simulation package (VASP)[52–55] are used in this paper. In order to incorporate the exchange-correlation from electrons, we use the local density approximation (LDA), as formulated by Ceperley and Adler[56] in all the calculations. The semicore electrons 3s and 3p for Y and 5d for Bi are treated as the valence electrons. The convergence criterion for self-consistent field calculation of energy and band structure calculations was chosen as $10^{-8}$ eV, and the residual forces for relaxation calculations were minimized down to 5×$10^{-5}$ eV/Å per atom. The energy cutoff of plane wave basis was chosen as 400 eV for all calculations. A Monkhorst-Pack[56] k-mesh of 13×13×13 and 13×13×1 are used for bulk and slab calculations, respectively. To account for relativistic correction, we included spin-orbit coupling in all the simulations.

**Acknowledgements:**
Authors would like to thank the CRF and NRF IIT Delhi for providing characterization facilities. VB and AB acknowledge MHRD India for financial support. SS, NB and RC acknowledges the financial assistance received from SPARC proposal #754. NB and JS also acknowledge funding from the EPSRC through EP/S016430/1. We also thank Prof. Umesh V. Waghmare from JNCASR for useful suggestions in DFT calculations. The DFT results presented in the paper are based on the computations using the High Performance Computing cluster, Padum, at IIT Delhi.

**Author contributions:**
VB performed the experimental work and has done the complete analysis of results. AB performed the theoretical simulations and DFT calculations supervised by BKM. SS performed mK resistivity measurement with the help of JS and NB. VVK, RC and NB wrote the manuscript with inputs from all the authors. RC conceived the idea and supervised the work.

**Competing Interests:** The authors declare that they have no competing interests.

**References:**

1. Hasan, M. Z. & Kane, C. L. Colloquium: topological insulators. *Rev. Mod. Phys.* **82**, 3045 (2010).
2. Qi, X.-L. & Zhang, S.-C. Topological insulators and superconductors. *Rev. Mod. Phys.* **83**, 1057 (2011).
3. Burkov, A. A. Topological semimetals. *Nat. Mater.* **15**, 1145–1148 (2016).
4. Dirac, P. A. M. The quantum theory of the electron. *Proc. R. Soc. London. Ser. A, Contain. Pap. a Math. Phys. Character* **117**, 610–624 (1928).
5. Weyl, H. Elektron und gravitation. I. *Zeitschrift für Phys.* **56**, 330–352 (1929).
6. Hu, J., Xu, S.-Y., Ni, N. & Mao, Z. Transport of topological semimetals. *Annu. Rev. Mater. Res.* **49**, 207–252 (2019).
7. Lin, H. *et al.* Half-Heusler ternary compounds as new multifunctional experimental platforms for topological quantum phenomena. *Nat. Mater.* **9**, 546–549 (2010).
8. Al-Sawai, W. *et al.* Topological electronic structure in half-Heusler topological insulators. *Phys. Rev. B - Condens. Matter Mater. Phys.* **82**, 1–5 (2010).
9. Chadov, S. *et al.* Tunable multifunctional topological insulators in ternary Heusler compounds. *Nat. Mater.* **9**, 541–545 (2010).
10. Jia, Z.-Y. *et al.* Direct visualization of a two-dimensional topological insulator in the single-layer 1 T′− WT e 2. *Phys. Rev. B* **96**, 41108 (2017).
11. Peng, L. *et al.* Observation of topological states residing at step edges of WTe 2. *Nat. Commun.* **8**, 1–7 (2017).
12. Liu, Z. K. *et al.* Observation of unusual topological surface states in half-Heusler compounds LnPtBi (Ln= Lu, Y). *Nat. Commun.* **7**, 1–7 (2016).
13. Liu, C. *et al.* Metallic surface electronic state in half-Heusler compounds R PtBi (R= Lu, Dy, Gd). *Phys. Rev. B* **83**, 205133 (2011).
14. Bhardwaj, V., Bhattacharya, A., Nigam, A. K., Dash, S. P. & Chatterjee, R. Observation of surface dominated topological transport in strained semimetallic ErPdBi thin films. *Appl. Phys. Lett.* **117**, 132406 (2020).
15. Bhardwaj, V., Bhattacharya, A., Varga, L. K., Ganguli, A. K. & Chatterjee, R. Thickness dependent magneto-transport properties of topologically nontrivial DyPdBi thin films. *Nanotechnology* (2020). doi:10.1088/1361-6528/ab99f3
16. Bhardwaj, V. & Chatterjee, R. Topological Materials: New Quantum Phases of Matter. *Resonance* **25**, 431–441 (2020).
17. Bhardwaj, V. *et al.* Weak antilocalization and quantum oscillations of surface states in topologically nontrivial dypdbi (110) half heusler alloy. *Sci. Rep.* **8**,

1–9 (2018).
18. Pavlosiuk, O., Kaczorowski, D. & Wiśniewski, P. Superconductivity and Shubnikov–de Haas oscillations in the noncentrosymmetric half-Heusler compound YPtBi. *Phys. Rev. B* **94**, 35130 (2016).
19. Ando, Y. Topological insulator materials. *J. Phys. Soc. Japan* **82**, 102001 (2013).
20. Butch, N. P., Syers, P., Kirshenbaum, K., Hope, A. P. & Paglione, J. Superconductivity in the topological semimetal YPtBi. *Phys. Rev. B* **84**, 220504 (2011).
21. Nowak, B. & Kaczorowski, D. NMR as a probe of band inversion in topologically nontrivial half-Heusler compounds. *J. Phys. Chem. C* **118**, 18021–18026 (2014).
22. Shi, C. *et al.* NMR investigation of atomic and electronic structures of half-Heusler topologically nontrivial semimetals. *Phys. status solidi* **252**, 357–360 (2015).
23. Nowak, B., Pavlosiuk, O. & Kaczorowski, D. Band inversion in topologically nontrivial half-Heusler bismuthides: 209Bi NMR study. *J. Phys. Chem. C* **119**, 2770–2774 (2015).
24. Zhang, X. *et al.* NMR evidence for the topologically nontrivial nature in a family of half-Heusler compounds. *Sci. Rep.* **6**, 23172 (2016).
25. Souza, J. C., Lesseux, G. G., Urbano, R. R., Rettori, C. & Pagliuso, P. G. Diffusive-like effects and possible non trivial local topology on the half-Heusler YPdBi compound. *AIP Adv.* **8**, 55713 (2018).
26. Souza, J. C. *et al.* Crystalline electric field study in a putative topologically trivial rare-earth doped YPdBi compound. *J. Phys. Condens. Matter* **31**, 465701 (2019).
27. Wang, W. *et al.* Large linear magnetoresistance and Shubnikov-de Hass oscillations in single crystals of YPdBi Heusler topological insulators. *Sci. Rep.* **3**, 2181 (2013).
28. Haase, M. G., Schmidt, T., Richter, C. G., Block, H. & Jeitschko, W. Equiatomic rare earth (Ln) transition metal antimonides LnTSb (T= Rh, lr) and bismuthides LnTBi (T= Rh, Ni, Pd, Pt). *J. Solid State Chem.* **168**, 18–27 (2002).
29. Taskin, A. A., Sasaki, S., Segawa, K. & Ando, Y. Achieving surface quantum oscillations in topological insulator thin films of Bi2Se3. *Adv. Mater.* **24**, 5581–5585 (2012).
30. Taskin, A. A., Sasaki, S., Segawa, K. & Ando, Y. Manifestation of topological protection in transport properties of epitaxial Bi 2 Se 3 thin films. *Phys. Rev. Lett.* **109**, 66803 (2012).
31. Bao, L. *et al.* Weak anti-localization and quantum oscillations of surface

states in topological insulator Bi 2 Se 2 Te. *Sci. Rep.* **2**, 726 (2012).
32. Xu, G. *et al.* Weak antilocalization effect and noncentrosymmetric superconductivity in a topologically nontrivial semimetal LuPdBi. *Sci. Rep.* **4**, 5709 (2014).
33. Pavlosiuk, O., Kaczorowski, D. & Wiśniewski, P. Shubnikov-de Haas oscillations, weak antilocalization effect and large linear magnetoresistance in the putative topological superconductor LuPdBi. *Sci. Rep.* **5**, 9158 (2015).
34. Hikami, S., Larkin, A. I. & Nagaoka, Y. Spin-orbit interaction and magnetoresistance in the two dimensional random system. *Prog. Theor. Phys.* **63**, 707–710 (1980).
35. Chen, J. *et al.* Tunable surface conductivity in Bi 2 Se 3 revealed in diffusive electron transport. *Phys. Rev. B* **83**, 241304 (2011).
36. Steinberg, H., Laloë, J.-B., Fatemi, V., Moodera, J. S. & Jarillo-Herrero, P. Electrically tunable surface-to-bulk coherent coupling in topological insulator thin films. *Phys. Rev. B* **84**, 233101 (2011).
37. Hirschberger, M. *et al.* The chiral anomaly and thermopower of Weyl fermions in the half-Heusler GdPtBi. *Nat. Mater.* **15**, 1161–1165 (2016).
38. Pavlosiuk, O., Kaczorowski, D. & Wiśniewski, P. Superconductivity and Shubnikov-de Haas oscillations in the noncentrosymmetric half-Heusler compound YPtBi. *Phys. Rev. B* **94**, 1–7 (2016).
39. Pavlosiuk, O., Kaczorowski, D., Fabreges, X., Gukasov, A. & Wiśniewski, P. Antiferromagnetism and superconductivity in the half-Heusler semimetal HoPdBi. *Sci. Rep.* **6**, 18797 (2016).
40. Shoenberg, D. Magnetic Oscillations in Metals Cambridge Univ. *Press. Cambridge* (1984).
41. Lifshitz, I. M. & Kosevich, A. M. Theory of magnetic susceptibility in metals at low temperatures. *Sov. Phys. JETP* **2**, 636–645 (1956).
42. Mikitik, G. P. & Sharlai, Y. V. Manifestation of berry's phase in metal physics. *Phys. Rev. Lett.* **82**, 2147–2150 (1999).
43. Xiao, D., Chang, M.-C. & Niu, Q. Berry phase effects on electronic properties. *Rev. Mod. Phys.* **82**, 1959 (2010).
44. Zhang, Y., Tan, Y.-W., Stormer, H. L. & Kim, P. Experimental observation of the quantum Hall effect and Berry's phase in graphene. *Nature* **438**, 201–204 (2005).
45. He, L. *et al.* Accuracy of generalized gradient approximation functionals for density-functional perturbation theory calculations. *Phys. Rev. B* **89**, 64305 (2014).
46. Fu, L., Kane, C. L. & Mele, E. J. Topological insulators in three dimensions. *Phys. Rev. Lett.* **98**, 106803 (2007).
47. Nakajima, Y. *et al.* Topological RPdBi half-Heusler semimetals: A new

family of noncentrosymmetric magnetic superconductors. *Sci. Adv.* **1**, e1500242 (2015).
48. Radmanesh, S. M. A. *et al.* Evidence for unconventional superconductivity in half-Heusler YPdBi and TbPdBi compounds revealed by London penetration depth measurements. *Phys. Rev. B* **98**, 241111 (2018).
49. Meinert, M. Unconventional superconductivity in YPtBi and related topological semimetals. *Phys. Rev. Lett.* **116**, 137001 (2016).
50. Blöchl, P. E. Projector augmented-wave method. *Phys. Rev. B* **50**, 17953 (1994).
51. Kresse, G. & Joubert, D. From ultrasoft pseudopotentials to the projector augmented-wave method. *Phys. Rev. b* **59**, 1758 (1999).
52. Kresse, G. & Hafner, J. Ab initio molecular dynamics for liquid metals. *Phys. Rev. B* **47**, 558 (1993).
53. Kresse, G. & Furthmüller, J. Efficient iterative schemes for ab initio total-energy calculations using a plane-wave basis set. *Phys. Rev. B* **54**, 11169 (1996).
54. Kresse, G. & Furthmüller, J. Efficiency of ab-initio total energy calculations for metals and semiconductors using a plane-wave basis set. *Comput. Mater. Sci.* **6**, 15–50 (1996).
55. Kresse, G. & Hafner, J. Ab initio molecular-dynamics simulation of the liquid-metal–amorphous-semiconductor transition in germanium. *Phys. Rev. B* **49**, 14251 (1994).
56. Ceperley, D. M. & Alder, B. J. Ground state of the electron gas by a stochastic method. *Phys. Rev. Lett.* **45**, 566 (1980).